\documentclass[conference]{IEEEtran}
\IEEEoverridecommandlockouts
\usepackage{graphicx}
\usepackage{paralist}
\usepackage{algorithm}
\usepackage{algorithmicx}
\usepackage{algpseudocode}
\usepackage[dvipsnames]{xcolor}
\usepackage{amsmath}
\usepackage{hyperref}
\usepackage{caption}
\usepackage{subcaption}
\usepackage{ulem}
\usepackage{xspace}
\usepackage{tabularx}
\usepackage{booktabs}
\usepackage{multirow}

\hypersetup{
    colorlinks = true,
    linkcolor=red,
    citecolor=red,
    urlcolor=blue,
    linktocpage % so that page numbers are clickable in toc
}

\algblock{Input}{EndInput}
\algnotext{EndInput}
\algblock{Output}{EndOutput}
\algnotext{EndOutput}
\newcommand{\Desc}[2]{\State \makebox[2em][l]{#1}#2}

\newcommand{\simgrid}{SimGrid\xspace}
\newcommand{\wrench}{WRENCH\xspace}

\begin{document}

\title{Modeling the Linux page cache for accurate simulation of data-intensive applications}

\author{
  \IEEEauthorblockN{
    Hoang-Dung Do\IEEEauthorrefmark{1},
    Val\'erie Hayot-Sasson\IEEEauthorrefmark{1},
    Rafael Ferreira da Silva\IEEEauthorrefmark{3},
    Christopher Steele\IEEEauthorrefmark{4},
    Henri Casanova\IEEEauthorrefmark{2},
    Tristan Glatard\IEEEauthorrefmark{1}
  }\\
  \IEEEauthorblockA{
    \IEEEauthorrefmark{1}Department of Computer Science and Software Engineering, Concordia University, Montreal, Canada\\
    \IEEEauthorrefmark{2}Department of Information and Computer Sciences, University of Hawai`i at M\=anoa, USA\\
    \IEEEauthorrefmark{3}Information Sciences Institute, University of Southern California, Marina Del Rey, CA, USA\\
    \IEEEauthorrefmark{4}Department of Psychology, Concordia University, Montreal, Canada
  }
}

\maketitle

    \begin{abstract}

    The emergence of Big Data in recent years has resulted in a growing
    need for efficient data processing solutions. While infrastructures
    with sufficient compute power are available,
    the I/O bottleneck remains. The Linux page cache is an efficient
    approach to reduce I/O overheads, but few
    experimental studies of its interactions with Big Data applications exist,
    partly due to limitations of
    real-world experiments. Simulation is a popular approach to address
    these issues, however, existing simulation frameworks do not simulate
    page caching fully, or even at all.  As a result, simulation-based
    performance studies of data-intensive applications lead to inaccurate
    results.

    In this paper, we propose an I/O simulation model that includes
    the key features of the Linux page cache. We have implemented this model
    as part of the \wrench workflow simulation framework, which itself
    builds on the popular \simgrid distributed systems simulation
    framework. Our model and its implementation enable the simulation
    of both single-threaded and multithreaded applications, and of both
    writeback and writethrough caches for local or network-based
    filesystems. We evaluate the accuracy of our model in different
    conditions, including sequential and concurrent applications, as
    well as local and remote I/Os. We find that our page cache model
    reduces the simulation error by up to an order of magnitude when
    compared to state-of-the-art, cacheless simulations.

    \end{abstract}

    \section{Introduction}

        The Linux page cache plays an important role in
        reducing filesystem data transfer times. With the page cache, previously
        read data can be re-read directly from memory, and written data can be written to
        memory before being asynchronously flushed to disk, resulting in improved I/O performance
        on slower storage devices. The performance improvements
        depend on many factors including the total amount of memory,
        the amount of data being written (i.e., dirty data), and the amount of memory available for
        written data. All these factors are important when determining the impact of I/O on
        application performance, particularly in data-intensive applications.

        % Swapped the order the two paragraphs and merged them into one
        The number of data-intensive applications has been steadily rising as a result of
        open-data and data sharing initiatives. Due to the sheer size of the data being
        processed, these applications must be executed on large-scale infrastructures
    such as High Performance Computing (HPC) clusters or the cloud.  It
    is thus crucial to quantify the performance of these applications
    on these platforms. The goals include determining which type of hardware/software
    stacks are best suited to different application classes, as well as
    understanding the limitations of current algorithms, designs and
    technologies. Unfortunately, performance
        studies relying on real-world experiments on compute platforms
        face several difficulties (high operational costs, labor-intensive experimental setups,
        shared platforms with dynamic loads that hinder reproducibility of results) and shortcomings
        (experiments are limited to the available platform/software configurations, which precludes
        the exploration of hypothetical scenarios). 
    Simulations address these concerns by providing models
    and abstractions for the performance of computer hardware, such as
    CPU, network and storage. As a result, simulations provide a
    cost-effective, fast, easy and reproducible way to evaluate
    application performance on arbitrary platform configurations. It thus comes
        as no surprise that a large number of simulation frameworks have been developed and used 
        for research and development~\cite{ optorsim, gridsim, groudsim, cloudsim, nunez2012simcan,nunez2012icancloud, mdcsim, dissect_cf, cloudnetsimplusplus, fognetsimplusplus, casanova2014simgrid, ROSS, casanova2020fgcs}.

        Page caching is an ubiquitous technique for mitigating the I/O bottleneck.
        As such, it is necessary to model it
        when simulating data-intensive applications.
        While existing simulation frameworks of parallel and distributed computing
        systems  capture many relevant features of hardware/software stacks, 
        they lack the ability to simulate page cache with enough details to capture key features such
        as dirty data and cache eviction policies~\cite{nunez2012simcan,nunez2012icancloud}. 
        Some simulators, such as the one in~\cite{xu2018saving}, do capture such features,
        but are domain-specific. 

        %However, there is a trade-off between accuracy and scalability, which has been
        %indicated in some reviews \cite{casanova2014simgrid} \cite{byrne2017review}.
        %This means that in most cicumstances, users have to, more or less, sacrifice
        %accuracy for the scalability and performance when their platforms grow to
        %hundreds or thousands nodes.

    In this work, we present \wrench-cache, a page cache simulation model
    implemented in \wrench~\cite{casanova2020fgcs}, a workflow simulation
    framework based on the popular \simgrid distributed simulation
    toolkit~\cite{casanova2014simgrid}. Our contributions are:
    \begin{compactitem}
        \item A page cache simulation model that supports 
    both single-threaded and multithreaded applications, and both
    writeback and writethrough caches for local or network-based
    filesystems;
        \item An implementation of this model in \wrench; 
        \item An evaluation of the accuracy and scalability of our model, and of its implementation,
              for multiple applications, execution scenarios, and page cache configurations. 
    \end{compactitem}

    \section{Related Work}
    \label{relatedwork}

        \subsection{Page cache}

        %To improve performance of disk accesses, the page cache was introduced to the Linux kernel. %It contains pages referring to physical pages on disk \cite{linuxdev3rd2010}
        Page cache offsets the cost of disk I/O by enabling I/O to occur directly from memory.
        When a file is first loaded into memory, the file is read from disk and loaded into the page cache as
        a series of pages. Subsequent reads to any of the file pages located in memory will result in
        a \textit{cache hit}, meaning the I/O can occur directly from without disk involvement.
        Any accessed page not loaded in memory results in a \textit{cache miss}, resulting in
        the page being read directly from disk.
        
        Written pages can also contribute to future application cache hits. When use of page cache
        is enabled for a given filesystem, through the enabling of writeback or writethrough cache,
        all written pages are written first to page cache, prior to being written to disk.
        Accessing of these written pages may result in cache hits, should the pages remain in memory.

        The kernel may also provide improved write performance if writeback cache is enabled. Unlike writethrough,
        where the data is synchronously written from memory to disk, writeback enables asynchronous writes.
        With writeback, the application may proceed with execution once the data has been
        written to memory, even if it has not yet been materialized to disk.  
        The writeback strategy is considered to outperform writethrough as well as
        direct I/O (page cache bypassed for I/O) as it delays disk writes to perform a bulk write at a later time
        \cite{linuxdev3rd2010}.

        Cache eviction and flushing strategies are integral to proper page cache functioning.
        Whenever space in memory becomes limited, either as a result of application memory
        or page cache use, page cache data may be evicted. Only data that
        has been persisted to storage (clean pages) can be flagged for eviction and removed from
        memory. Written data that has not yet been persisted to disk (dirty data) must first
        be copied (flushed) to storage prior to eviction. When sufficient memory is
        being occupied, the flushing process is synchronous. However, even when
        there is sufficient available memory, written data will be flushed to disk
        at a predefined interval through a process known as \textit{periodical flushing}.
        Periodical flushing only flushes expired dirty pages, which remain dirty in
        page cache longer than an expiration time configured in the kernel.
        Different cache eviction algorithms have also been proposed
        \cite{owda2014comparison}.

        The Linux kernel uses a two-list strategy to flag pages for eviction.
        The two-list strategy is based on a least recently used (LRU) policy
        and uses an active and inactive list in its implementation.
        If accessed pages are not in the page cache, they are added to the inactive list.
        Should pages located on the inactive list be accessed, they will be moved from
        the inactive to the active list.
        The lists are also kept balanced by moving pages from the active list
        to the inactive list when the active list grows too large.
        Thus, the active list only contains pages which are accessed more than once
        and not evictable, while the inactive list includes pages accessed once only,
        or pages that have been accessed more than once but moved from the active list.
        Both lists operate using LRU eviction policies, meaning that data that has
        not be accessed recently will be moved first.

        \subsection{Simulation}

        Many simulation frameworks have been developed to enable the
        simulation of parallel and distributed
        applications~\cite{optorsim, gridsim, groudsim, cloudsim,
        nunez2012simcan,nunez2012icancloud, mdcsim, dissect_cf,
        cloudnetsimplusplus, fognetsimplusplus, casanova2014simgrid,
        ROSS, casanova2020fgcs}. These frameworks implement simulation
        models and abstractions to aid the development of simulators
        for studying the functional and performance behaviors of
        application workloads executed on various hardware/software
        infrastructures. 

        The two main concerns for simulation are accuracy,
        the ability to faithfully reproduce real-world executions, and
        scalability, the ability to simulate large/long real-world
        executions quickly and with low RAM footprint. The above
        frameworks achieve different compromises between the two.  At
        one extreme are discrete-event models that capture
        ``microscopic'' behaviors of hardware/software systems (e.g.,
        packet-level network simulation, block-level disk simulation,
        cycle-accurate CPU simulation), which favor accuracy over
        speed.  At the other extreme are analytical models that capture
        ``macroscopic'' behaviors via mathematical models.  While these
        models lead to fast simulation, they must be developed
        carefully if high levels of accuracy are to be
        achieved~\cite{velhoTOMACS2013}. 

        In this  work, we use the \simgrid and \wrench simulation
        frameworks.  The years of research and development invested in
        the popular \simgrid simulation framework~\cite{casanova2014simgrid}, have
        culminated in a set of state-of-the-art macroscopic simulation
        models that yield high accuracy, as demonstrated by
        (in)validation studies and comparisons to competing
        frameworks~\cite{smpi_validity, velhoTOMACS2013, simutool_09,
        nstools_07, lebre2015, pouilloux:hal-01197274,
        smpi_tpds2017,  7885814, 8048921, 7384330}.  But one
        significant drawback of \simgrid is that its simulation
        abstractions are low-level, meaning that implementing a
        simulator of complex systems can be
        labor-intensive~\cite{kecskemeti_2014}. To remedy this problem,
        the \wrench simulation framework~\cite{casanova2020fgcs}
        builds on top of \simgrid to provide higher-level simulation
        abstractions, so that simulators of complex applications and
        systems can be implemented with a few hundred lines.

        Although the Linux page cache has a large impact on I/O
        performance, and thus on the execution of data-intensive
        applications, its simulation is rarely considered in the above
        frameworks.  Most frameworks merely simulate I/O operations
        based on storage bandwidths and capacities.  The SIMCAN
        framework does models page caching by storing data accessed on
        disk in a block cache~\cite{nunez2012simcan}.  Page cache is
        also modeled in iCanCloud through a component that manages
        memory accesses and cached data~\cite{nunez2012icancloud}.
        However, the scalability of the iCanCloud simulator is limited
        as it uses microscopic models.  Besides, none
        of these simulators provide any writeback cache simulator nor
        cache eviction policies through LRU lists.  Although cache
        replacement policies are applied in~\cite{xu2018saving} to
        simulate in-memory caching, this simulator is specific to
        energy consumption of multi-tier heterogeneous networks.

        In this study, we implement a page cache simulation model in the
        \wrench framework. We targeted \wrench because it is a recent,
        actively developed framework that provides convenient simulation
        abstractions, because it is extensible, and because it reuses
        \simgrid's scalable and accurate models.

    \section{Methods}
    \label{method}
\begin{figure}
           \centering
           \includegraphics[width=0.85\columnwidth]{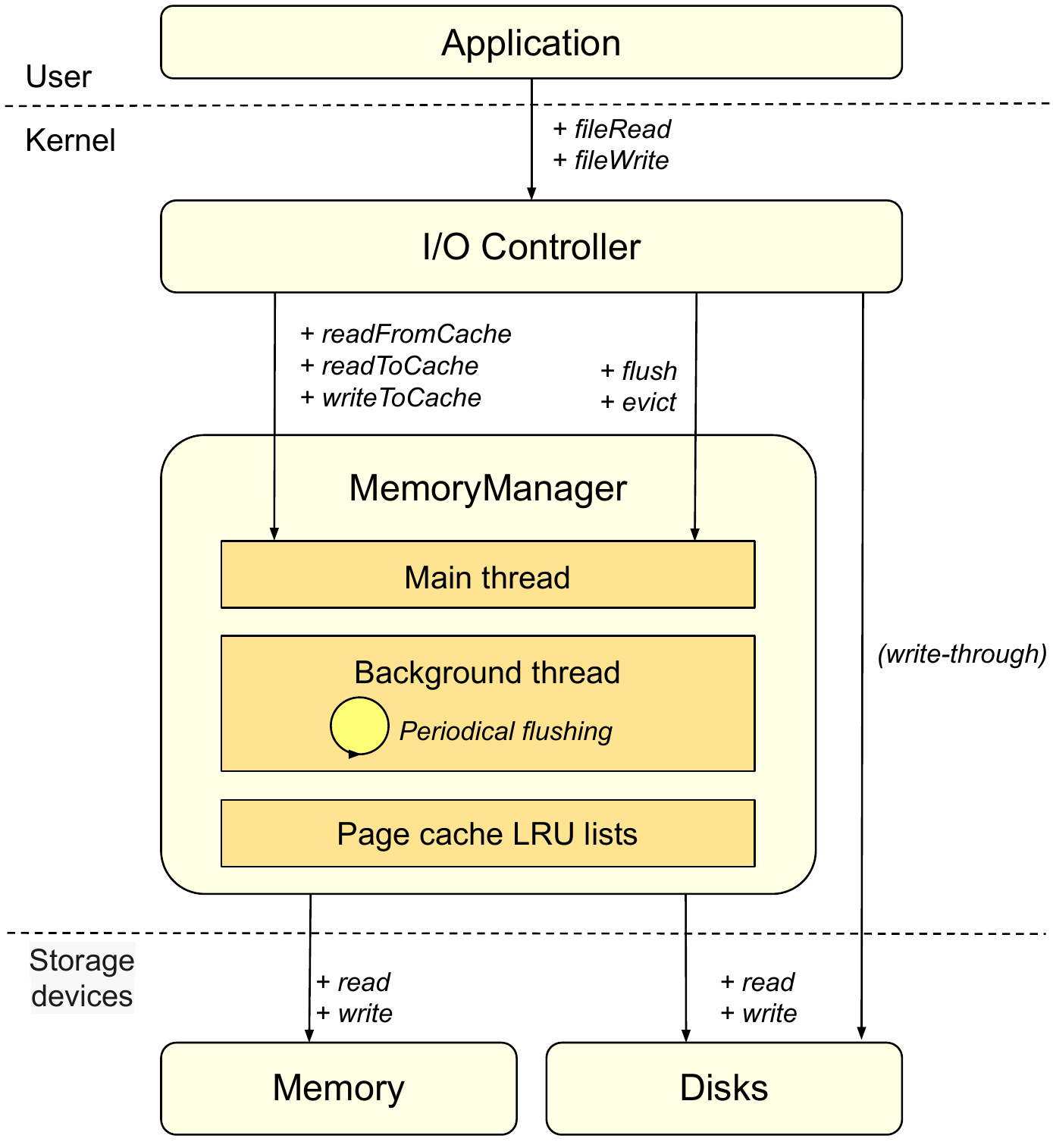}
           \caption{Overview of the page cache simulator.
           Applications send file read or write requests to the
           I/O Controller that orchestrates flushing, eviction, cache
           and disk accesses with the Memory Manager. Concurrent accesses to storage
           devices (memory and disk) are simulated using existing models.}
           \label{fig:interaction}
    \end{figure}

    We separate our simulation model in two components, the I/O
    Controller and the Memory Manager, which together simulate
    file reads and writes (Figure~\ref{fig:interaction}).
    To read or write a file chunk, a simulated application sends a
    request to the I/O Controller. The I/O Controller interacts as needed with
    the Memory Manager to free memory through flushing or eviction,
    and to read or write cached data. The Memory Manager
    implements these operations, simulates periodical flushing
    and eviction, and reads or writes to disk when necessary.
    In case the writethrough strategy is used, the I/O Controller directly writes to disk, 
    cache is flushed if needed and written data is added to page cache.

    \subsection{Memory Manager}

    The Memory Manager simulates two parallel threads: the main one
    implements flushing, eviction, and cached I/Os synchronously, whereas
    the second one, which operates in the background, periodically searches for
    expired dirty data in LRU lists and flushes this data to disk. We
    use existing storage simulation models~\cite{lebre2015} to simulate disk and
    memory, characterized by their storage capacity, read and write
    bandwidths, and latency. These models account for
    bandwidth sharing between concurrent memory or disk accesses.

    \subsubsection{Page cache LRU lists}

    In the Linux kernel, page cache LRU lists contain file pages. However,
    due to the large number of file pages, simulating lists of pages
    induces substantial overhead.
    Therefore, we introduce the concept of a data block as a unit to represent data
    cached in memory. A data block is a subset of file pages stored in
    page cache that were accessed in the same I/O operation.
    A data block stores the file name, block size, last access
    time, a dirty flag that represents whether the data is clean (0)
    or dirty (1), and an entry (creation) time.
    Blocks can have different sizes and a given file can have multiple
    data blocks in page cache. In addition, a data block can be split into an
    arbitrary number of smaller blocks.
    \begin{figure}
           \centering
           \includegraphics[width=\columnwidth]{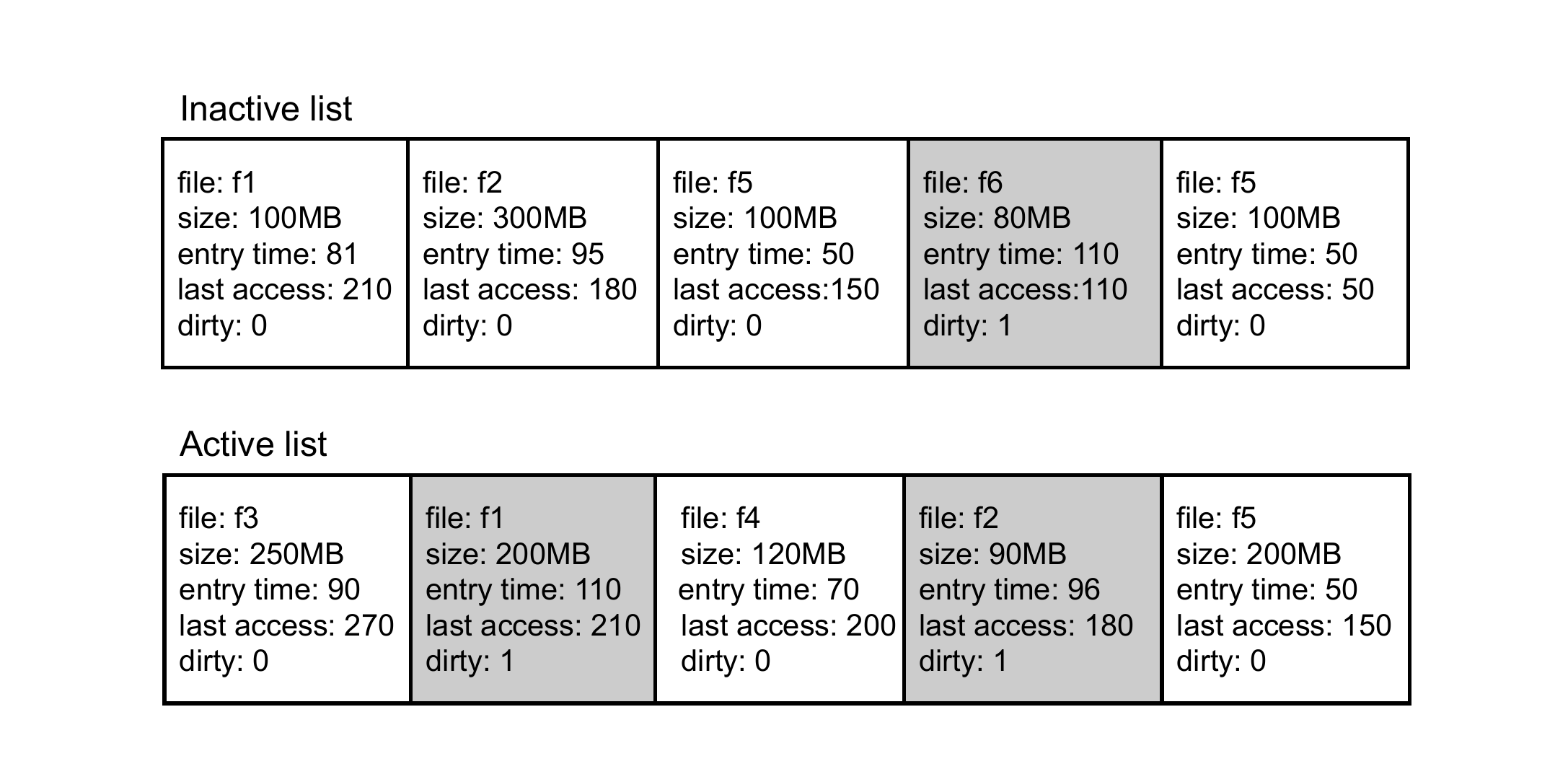}
           \caption{Model of page cache LRU lists with data blocks.}
           \label{fig:lrulist}
    \end{figure}

    We model page cache LRU lists as
    two lists of data blocks, an active list and an inactive list, both ordered by
    last access time (earliest first, Figure~\ref{fig:lrulist}).
    As in the kernel, our simulator limits the size of the active list to
    twice the size of the inactive list, by moving least recently
    used data blocks from the active list to the inactive list~\cite{gorman2004understanding, linuxdev3rd2010}.

    At any given time, a file can be partially cached, completely cached,
    or not cached at all. A cached data block can only reside in one of two
    LRU lists. The first time they are accessed, blocks are
    added to the inactive list. On subsequent accesses, blocks of the
    inactive list are moved to the top of the active list. Blocks
    written to cache are marked dirty until flushed.

    \subsubsection{Reads and writes}

\begin{figure}
           \centering
           \includegraphics[width=\columnwidth]{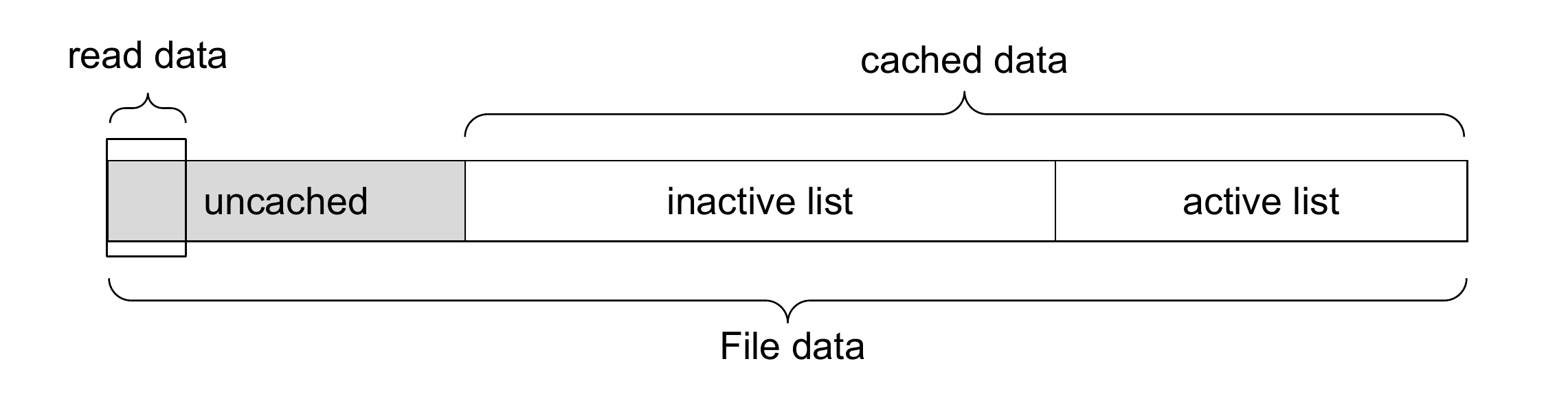}
           \caption{File data read order. Data is read from left to right: uncached data
           is read first, followed by data from the inactive list, and finally data from the active list. }
           \label{fig:read_order}
    \end{figure}

    Our simulation model supports chunk-by-chunk file accesses
    with a user-defined chunk size. However, for simplicity, we assume that file pages are
    accessed in a round-robin fashion rather than fully randomly.
    Therefore, when a file is read, cached data is read only after all uncached data was read, and data from the inactive list is read
    before data from the active list
    (data reads occur from left to right in Figure~\ref{fig:read_order}).
    When a chunk of \emph{uncached} data is read, a new clean block is created
    and appended to the inactive list.
    When a chunk of \emph{cached} data is read, one or more existing data blocks in the LRU lists are accessed.
    If these blocks are clean, we merge them together, update the access time and size of the resulting block,
    and append it to the active list.
    If the blocks are dirty, we move them independently to the active list, to preserve their entry time.
    Because the chunk and block sizes may be different, there are situations
    where a block is not entirely read.
    In this case, the block is split in two smaller blocks and one of them is re-accessed.
    % From Tristan: not sure where to put this nor if it's necessary:
    % , in which chunks are read/written
    % until file is entirely read/written.

    For file writes, we assume that all data to be written is
    uncached. Thus, each time a chunk is written, we create a block of dirty data
    and append it to the inactive list.

    \subsubsection{Flushing and eviction}

    The main simulated thread in the Memory Manager can flush or evict data from the
    memory cache. The data flushing simulation
    function takes the amount of data to flush as parameter. While
    this amount is not reached and dirty
    blocks remain in cache, this function traverses the sorted
    inactive list, then the sorted active list, and writes the
    least recently used dirty block to disk, having set its dirty
    flag to 0. In case the amount of data to flush requires that a
    block be partially flushed, the block is split in two blocks,
    one that is flushed and one that remains dirty. The time needed
    to flush data to disk is simulated by the storage model.

    The cache eviction simulation also runs in
    the main thread. It frees up the page cache by traversing and deleting
    least recently used clean data blocks in the inactive list.
    The amount of data to evict is passed as a parameter and data blocks are deleted
    from the inactive list until the evicted data reaches the required amount,
    or until there is no clean block left in the list.
    If the last evicted block does not have to be entirely evicted, the block is split in two blocks,
    and only one of them is evicted.
    The overhead of the cache eviction algorithm is not part of the simulated time
    since cache eviction time is negligible in real systems. % \tristan{a reference would be welcome}.

    \begin{algorithm}[b]\caption{Periodical flush simulation in Memory Manager}\label{alg:pdflush}
        \small
        \begin{algorithmic}[1]
            \Input
                \Desc{in}{page cache inactive list}
                \Desc{ac}{page cache active list}
                \Desc{t}{predefined flushing time interval}
                \Desc{exp}{predefined expiration time}
                \Desc{sm}{storage simulation model}
               \EndInput
               \While{host is on}
                \State blocks = expired\_blocks(exp, in) + expired\_blocks(exp, ac)
                \State flushing\_time = 0
                \For{blk in blocks} 
                  \State blk.dirty = 0 
                  \State flushing\_time = flushing\_time + sm.write(blocks)
                \EndFor
                \If{flushing\_time $<$ t}
                    \State sleep(t - flushing\_time)
                \EndIf  
            \EndWhile
        \end{algorithmic}
    \end{algorithm}

    Periodical flushing is simulated in the Memory Manager
    background thread. As in the Linux kernel, a dirty block
    in our model is considered expired if
    the duration since its entry time is longer than a
    predefined expiration time.
    Periodical flushing is simulated as an infinite loop in which
    the Memory Manager searches for dirty blocks and flushes them to disk (Algorithm~\ref{alg:pdflush}).
    % From Tristan: the algorithm is quite straightforward, I don't think this is
    % necessary
    % In each repetition, Memory Manager finds expired dirty blocks in two
    % page cache LRU lists (line 8), simulates writes of data of these blocks
    % to disk (line 9), and mark them as clean (line 10).
    % If the flushing time does not exceed our time interval, the thread is put
    % to sleep for the remaining time (lines 11-13).
    % Then based on the state of the host at current simulated time,
    % the algorithm continues or finishes the loop.
    Because periodical flushing is simulated as a background thread, it can happen concurrently
    with disk I/O initiated by the main thread. This is taken into account by the
    storage model and reflected in simulated I/O time.

    \subsection{I/O Controller}

    \begin{algorithm}\caption{File chunk read simulation in I/O Controller}
    \label{alg:read}
        \small
        \begin{algorithmic}[1]
            \Input
                \Desc{cs}{chunk size}
                \Desc{fn}{file name}
                \Desc{fs}{file size (assumed to fit in memory)}
                \Desc{mm}{MemoryManager object}
                \Desc{sm}{storage simulation model}
               \EndInput
               \State disk\_read = min(cs, fs - mm.cached(fn)) \Comment{To be read from disk}
               \State cache\_read = cs - disk\_read \Comment{To be read from cache}
               \State required\_mem = cs + disk\_read
               \State mm.flush(required\_mem - mm.free\_mem - mm.evictable, fn)
               \State mm.evict(required\_mem - mm.free\_mem, fn)
               \If {disk\_read $>$ 0}  \Comment{Read uncached data}
               \State sm.read(disk\_read)
               \State mm.add\_to\_cache(disk\_read, fn)
               \EndIf
               \If {cache\_read $>$ 0} \Comment{Read cached}
               \State mm.cache\_read(cache\_read)
            \EndIf
            \State mm.use\_anonymous\_mem(cs)
        \end{algorithmic}
    \end{algorithm}
    As mentioned previously, our model reads and writes file chunks in a
    round-robin fashion. To read a file chunk, simulated applications send
    chunk read requests to the I/O Controller which processes them using
    Algorithm~\ref{alg:read}. First, we calculate the amount of uncached
    data that needs to be read from disk, and the remaining amount is read
    from cache (line 7-8). The amount of memory required to read the chunk
    is calculated, corresponding to a copy of the chunk in anonymous memory
    and a copy of the chunk in cache (line 9).
    If there is not enough available memory, the Memory Manager is called
    to flush dirty data (line 10). If necessary, flushing is complemented by
    eviction (line 11). Note that, when called with negative arguments, functions
    \texttt{flush} and \texttt{evict} simply return and do not do anything. Then,
    if the block requires
    uncached data, the memory manager is called to read data from disk and to add this
    data to cache (line 14).
    If cached data needs to be read, the Memory Manager is called to simulate
    a cache read  and update the corresponding data blocks accordingly (line 17).
    Finally, the memory manager is called to deallocate the amount of anonymous memory used by the application (line 19).

    \begin{algorithm}\caption{File chunk write simulation in I/O Controller}
    \label{alg:write}
        \small
        \begin{algorithmic}[1]
            \Input
                \Desc{cs}{chunk size}
                \Desc{fn}{file name}
                \Desc{mm}{MemoryManager object}
               \EndInput
            \State remain\_dirty = dirty\_ratio * mm.avail\_mem - mm.dirty
            \If {remain\_dirty $>$ 0} \Comment{Write to memory}
                \State mm.evict(min(cs, remain\_dirty) - mm.free\_mem)
                \State mem\_amt = min(cs, mm.free\_mem)
                \State mm.write\_to\_cache(fn, mem\_amt)
            \EndIf
            \State remaining = cs - mem\_amt
            \While {remaining $>$ 0}  \Comment{Flush to disk, then write to cache}
                \State mm.flush(cs - mem\_amt)
                \State mm.evict(cs - mem\_amt  - mm.free\_mem)
                \State to\_cache = min(remaining, mm.free\_mem)
                \State mm.write\_to\_cache(fn, to\_cache)
                \State remaining = remaining - to\_cache
            \EndWhile

        \end{algorithmic}
    \end{algorithm}
    Algorithm~\ref{alg:write} describes our simulation of chunk writes in
    the I/O Controller.
    % Data is written to cache until the amount of dirty data in the cache
    % exceeds the dirty ratio times the amount of available memory. Data is then written to disk.
    Our algorithm initially checks the  amount of dirty data that
    can be written given the dirty ratio (line 5).
    If this amount is greater than 0, the Memory Manager is requested to evict
    data from cache if necessary (line 7).
    After eviction, the amount of data that can be written to
    page cache is calculated (line 8), and a cache write is simulated (line 9).
    If the dirty threshold is reached and there is still data to write,
    the remaining data is written to cache in a loop
    where we repeatedly flush and evict from the cache (line 12-18).

    The above model describes page cache in writeback
    mode. Our model also includes a write function in writethrough mode,
    which simply simulates a disk write with the amount of data passed in,
    then evicts cache if needed and adds the written data to the cache.

        \subsection{Implementation}

            We first created a standalone prototype
            simulator to evaluate the accuracy and correctness of our
            model in a simple scenario before integrating it in the more complex
            \wrench framework.
            The prototype uses the following basic storage model for
            both memory and disk:
            \begin{align*}
                & t_{r} = D / b_r \\
                & t_{w} = D / b_w\
            \end{align*}

            where:
            \begin{itemize}
                \item $t_{r}$ is the data read time
                \item $t_{w}$ is the data write time
                \item $D$ is the amount of data to read or write
                \item $b_r$ is the read bandwidth of the device
                \item $b_w$ is the write bandwidth of the device
            \end{itemize}

            This prototype does not simulate  bandwidth sharing and thus does not support
            concurrency: it is limited to single-threaded applications running on systems
            with a single-core CPU. We used this prototype for a first validation of our simulation
            model against a real sequential application running on a real system.
            The Python 3.7 source code is available at
            \url{https://github.com/big-data-lab-team/paper-io-simulation/tree/master/exp/pysim}.

            We also implemented our model as part of \wrench, enhancing its
            internal implementation and APIs with a page cache abstraction,
            and allowing users to activate the feature via a command-line
            argument. We used SimGrid's locking mechanism to handle
            concurrent accesses to page cache LRU lists by the two Memory
            Manager threads. For the experiments, we used
            \wrench 1.6 at commit
            \href{https://github.com/wrench-project/wrench/tree/67185374330d2c4bf274fce222c937e838df5b03}{6718537433},
            which uses \simgrid 3.25, available at
            \url{https://framagit.org/simgrid/simgrid}. Our implementation
            is now part of \wrench's master branch and will be available to
            users with the upcoming 1.8 release. \wrench provides a full \simgrid-based simulation 
            environment that supports, among other features, concurrent accesses to storage devices, 
            applications distributed on multiple hosts, network transfers, 
            and multi-threading. 

        \subsection{Experiments}

        Our experiments compared real executions with our Python prototype,
        with the original \wrench simulator, and with our \wrench-cache
        extension. Executions included single-threaded and multi-threaded
        applications, accessing data on local and network file systems. We
        used two applications: a synthetic one, created to evaluate the
        simulation model, and a real one, representative of neuroimaging
        data processing.

        Experiments were run on a dedicated cluster at
        Concordia University, with one login node, 9 compute nodes, and 4
        storage nodes connected with a 25 Gbps network. Each
        compute node had 2 $\times$ 16-core Intel(R) Xeon(R) Gold 6130 CPU
        @ 2.10GHz, 250~GiB of RAM, 6 $\times$ SSDs of 450~GiB each with the XFS
        file system, 378~GiB of tmpfs, 126~GiB of devtmpfs file system,
        CentOS~8.1 and NFS version 4. We used the \texttt{atop}
        and \texttt{collectl} tools to monitor and collect memory status
        and disk throughput. We cleared the page
        cache before each application run to ensure comparable
        conditions.

        \begin{table}[b]
            \centering
            \begin{tabularx}{0.8\columnwidth}{c>{\centering\arraybackslash}X}
            \toprule
                Input size (GB)  & CPU time (s)\\
            \midrule
                3      & 4.4 \\
                20  & 28 \\
                50  & 75 \\
                75  & 110 \\
                100  & 155 \\
            \bottomrule
            \end{tabularx}
            \caption{Synthetic application parameters}
            \label{table:cputime}
            \end{table}
        The synthetic application, implemented in C, consisted of three single-core,
        sequential tasks where each task read the file produced by the
        previous task, incremented every byte of this file to emulate real
        processing, and wrote the resulting data to disk. Files were
        numbered by ascending access times (File 1 was the file read by Task 1, etc).
         The anonymous memory used by the application
        was released after each task, which we also simulated in the Python
        prototype and in WRENCH-cache. As our focus was on I/O rather than compute, we measured
        application task CPU times on a cluster node
        (Table~\ref{table:cputime}), and used these durations in our
        simulations. For the Python prototype, we injected CPU times
        directly in the simulation. For \wrench and \wrench-cache, we
        determined the corresponding number of flops on a 1~Gflops CPU
        and used these values in the simulation. The simulated
        platform and application are available at
        commit \href{https://github.com/wrench-project/wrench/tree/ec6b43561b95977002258c0fe37a4ecad8f1d33f/examples/basic-examples/io-pagecache}{ec6b43561b}.

        We used the synthetic application in three experiments. In the
        first one (\textit{Exp~1}), we ran \emph{a single instance} of
        the application on a single cluster node, with different input file
        sizes (20~GB, 50~GB, 75~GB, 100~GB), and with all I/Os directed to
        the same local disk.
        In the second experiment (\textit{Exp~2}), we ran
        \emph{concurrent} instances of the application on a single node,
        all application instances operating on different files of size 3~GB
        stored in the same local disk. We varied the number of concurrent
        application instances from 1 to 32 since cluster nodes had 32 CPU
        cores.
        In the third experiment (\textit{Exp~3}), we used the same
        configuration as the previous one, albeit reading and writing
        on a 50-GiB NFS-mounted partition of a 450-GiB remote disk of
        another compute node. As is commonly configured in HPC
        environments to avoid data loss, there was no client write cache
        and the server cache was configured as writethrough instead of
        writeback. NFS client and server read caches were enabled. 
        Therefore, all the writes happened at disk bandwidth, but
        reads could benefit from cache hits.

        The real application was a workflow of the Nighres
        toolbox~\cite{huntenburg2018nighres}, implementing cortical
        reconstruction from brain images in four steps: skull stripping,
        tissue classification, region extraction, and cortical
        reconstruction. Each step read files produced by the previous step,
        and wrote files that were or were not read by the subsequent step.
        More information on this application is available in the Nighres
        documentation at
        \url{https://nighres.readthedocs.io/en/latest/auto_examples/example_02_cortical_depth_estimation.html}.
        The application is implemented as a Python script that calls Java
        image-processing routines. We used Python 3.6, Java 8, and Nighres
        1.3.0. We patched the application to remove lazy data loading and
        data compression, which made CPU time difficult to separate from
        I/O time, and to capture task CPU times to inject them in the
        simulation. The patched code is available at
        \url{https://github.com/dohoangdzung/nighres}.
        \begin{table}[t]
            \centering
            \begin{tabular}{lccc}
            \toprule
                \multicolumn{1}{c}{Workflow step}& Input size       & Output size      & CPU time\\
                                       & (MB)             & (MB)             & (s)\\
            \midrule
               Skull stripping         &  295             & 393               & 137 \\
               Tissue classification   &  197              & 1376              & 614 \\
               Region extraction       &  1376             & 885              & 76 \\
               Cortical reconstruction &  393              & 786              & 272\\
            \bottomrule
            \end{tabular} 
            \caption{Nighres application parameters}
            \label{table:nighres_stats}
            \end{table}
            \begin{table}[b]
                \centering
                \begin{tabularx}{\columnwidth}{ll
                >{\centering\arraybackslash}X
                >{\centering\arraybackslash}X
                >{\centering\arraybackslash}X}
                \toprule
                    \multicolumn{2}{c}{Bandwidths}  & Cluster (real) & Python prototype & \wrench simulator\\
                \midrule
                \multirow{2}{*}{Memory}      & read  & 6860 & 4812 & 4812\\
                                             & write & 2764 & 4812 & 4812\\
                \multirow{2}{*}{Local disk}  & read  & 510  & 465  & 465\\
                                             & write & 420  & 465  & 465\\
                \multirow{2}{*}{Remote disk} & read  & 515  & -    & 445\\
                                             & write & 375  & -    & 445\\
                \multicolumn{2}{l}{Network}  & 3000  & -    & 3000\\
                \bottomrule
                \end{tabularx}
                \caption{Bandwidth benchmarks (MBps) and simulator configurations.
                The bandwidths used in the simulations were the average of the measured read and write bandwidths.
                Network accesses were not simulated in the Python prototype.}
                \label{table:benchmark}
                \end{table}

        We used the real application in the fourth experiment
        (\textit{Exp~4}), run on a single cluster node 
        using a single local disk. We processed data from
        participant 0027430 in the dataset of the Max Planck Institute for
        Human Cognitive and Brain Sciences available at
        \url{http://dx.doi.org/10.15387/fcp_indi.corr.mpg1}, leading to the
        parameters in Table~\ref{table:nighres_stats}.

        To parameterize the simulators, we benchmarked the
        memory, local disk, remote disk (NFS), and network bandwidths
        (Table~\ref{table:benchmark}). Since \simgrid, and thus \wrench, currently only supports
        symmetrical bandwidths, we use the mean of the read and write
        bandwidth values in our experiments.

            % Finally, because \wrench simulates applications base on network-communication,
            % we use an infinite bandwidth to eliminate network latency for local I/Os.

    \section{Results}
    \label{results}

    \begin{figure*}
        \centering
        \begin{subfigure}{\linewidth}
            \centering
               \includegraphics[width=\linewidth]{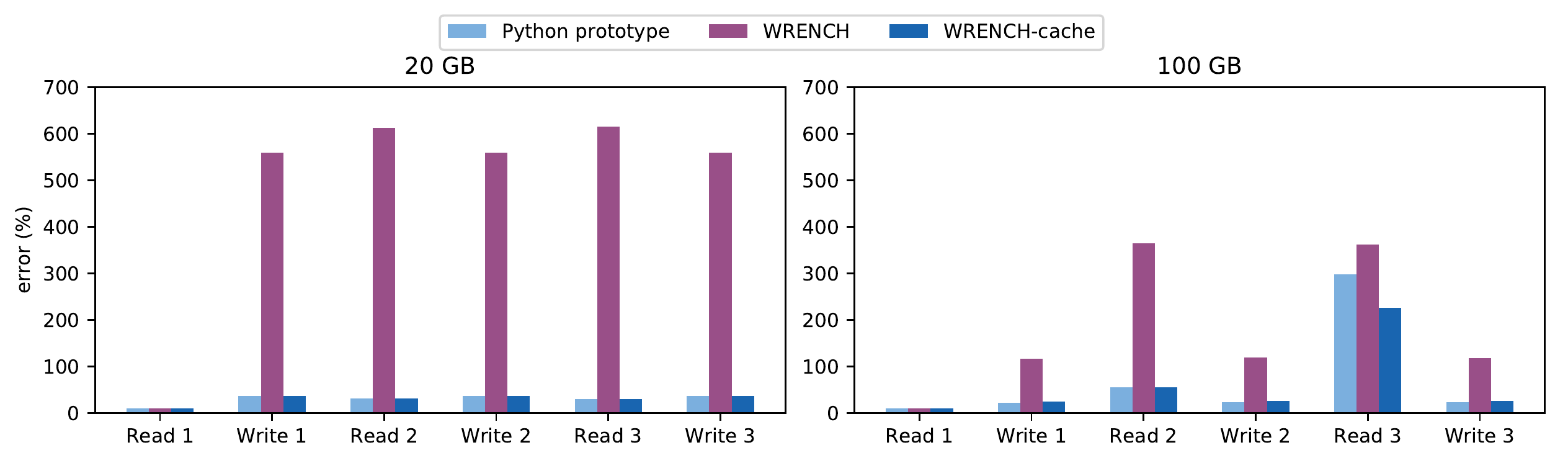}
               \vspace*{-0.7cm}
               \caption{Absolute relative simulation errors}
               \vspace*{0.5cm}
               \label{fig:single_error}
            \end{subfigure}
        \begin{subfigure}{\linewidth}
            \centering
            %    Gray shades represent task phases (read, compute and write).
            %    Lines represent memory usage along pipeline execution time.}
               \includegraphics[width=\linewidth]{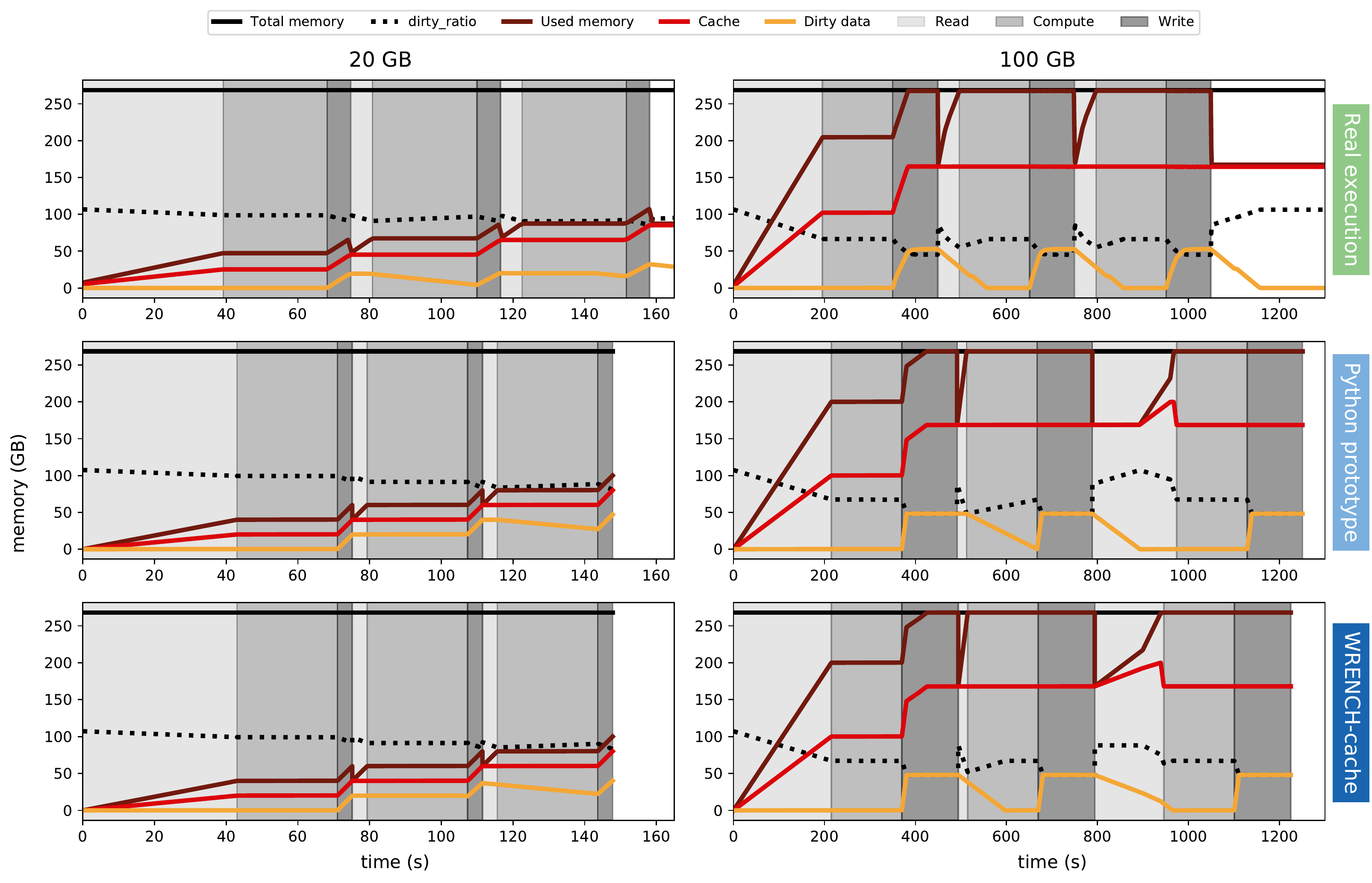}
               \vspace*{-0.7cm}
               \caption{Memory profiles}
               \vspace*{0.5cm}
               \label{fig:single_memprof}
        \end{subfigure}
        \begin{subfigure}{\linewidth}
            \centering
               \includegraphics[width=1.05\linewidth]{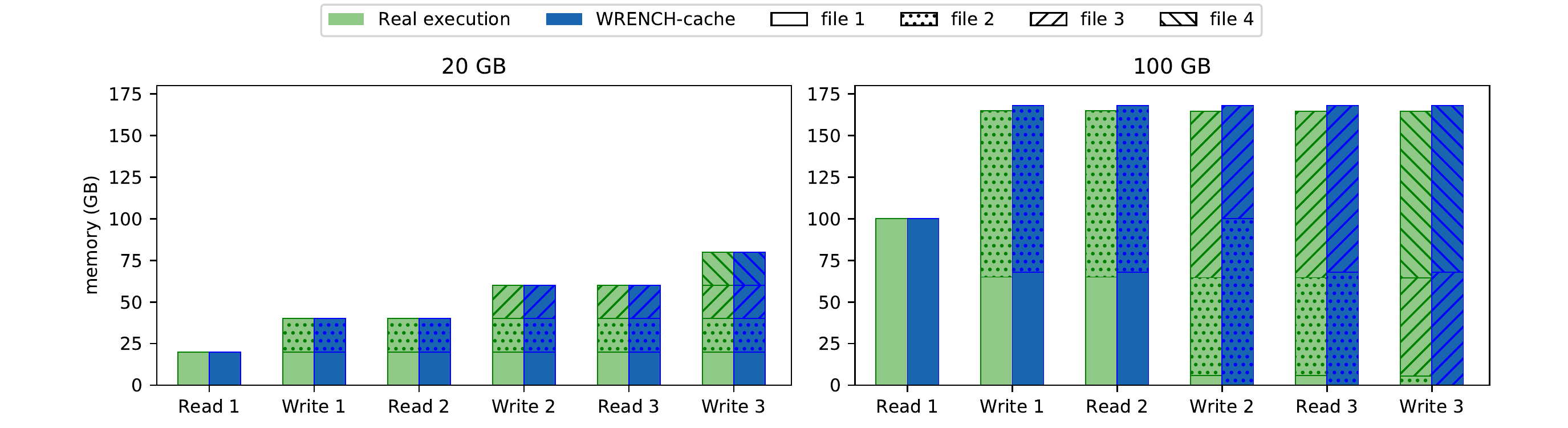}
               \caption{Cache contents \emph{after} application I/O operations}
            %    \textcolor{red}{Update real results of 20GB}}
               \label{fig:single_cache}
        \end{subfigure}
        \caption{Single-threaded results (\textit{Exp 1})}
        \end{figure*}

        \subsection{Single-threaded execution (Exp 1)}

        The page cache simulation model drastically reduced I/O simulation
        errors in each application task (Figure~\ref{fig:single_error}). The first read was not impacted
        as it only involved uncached data. Errors were reduced from an average
        of 345\% in the original \wrench to 46\% in the Python prototype and
        39\% in \wrench-cache. Unsurprisingly, the original \wrench simulator
        significantly overestimated read and write times, due to the lack
        of page cache simulation. Results with files of 50~GB and 75~GB
        showed similar behaviors and are not reported for brevity.

        \wrench simulation errors were substantially lower with 100~GB
        files than with 20~GB files, due to the fact that part of the
        100~GB file needed to be read and written to disk, the only storage
        device in \wrench, as it did not fit in cache. Conversely,
        simulation errors of the Python prototype and \wrench-cache were higher with
        100~GB files than with 20~GB files, due to idiosyncrasies in the kernel
        flushing and eviction strategies that could not be easily modeled.

        Simulated memory profiles were highly consistent with the real ones
        (Figure~\ref{fig:single_memprof}). With 20~GB files, memory profiles almost exactly matched the
        real ones, although dirty data seemed to be flushing faster in real
        life than in simulation, a behavior also
        observed with 100~GB files. With 100~GB files, used memory reached
        total memory during the first write, triggering dirty data
        flushing, and droped back to cached memory when application tasks
        released anonymous memory. Simulated cached memory was highly
        consistent with real values, except toward the end of Read 3 where
        it slightly increased in simulation but not in reality. This
        occurred due to the fact that after Write 2, File 3 was only partially
        cached in simulation whereas it was entirely cached in the real
        system. In all cases, dirty data remained under the dirty ratio as
        expected. The Python prototype and \wrench-cache exhibited nearly
        identical memory profiles, which reinforces the confidence in our
        implementations.

        The content of the simulated memory cache was also highly
        consistent with reality (Figure~\ref{fig:single_cache}). With 20~GB
        files, the simulated cache content exactly matched reality, since
        all files fitted in page cache. With 100~GB files, a slight
        discrepancy was observed after Write 2, which explains the
        simulation error previously mentioned in Read 3. In the real
        execution indeed, File 3 was entirely cached after Write 2, whereas
        in the simulated execution, only a part of it was cached. This was
        due to the fact that the Linux kernel tends to not evict pages that
        belong to files being currently written (File 3 in this case),
        which we could not easily reproduce in our model.

            \begin{figure*}
                \begin{subfigure}{\linewidth}
                    \centering
                    \includegraphics[width=\linewidth]{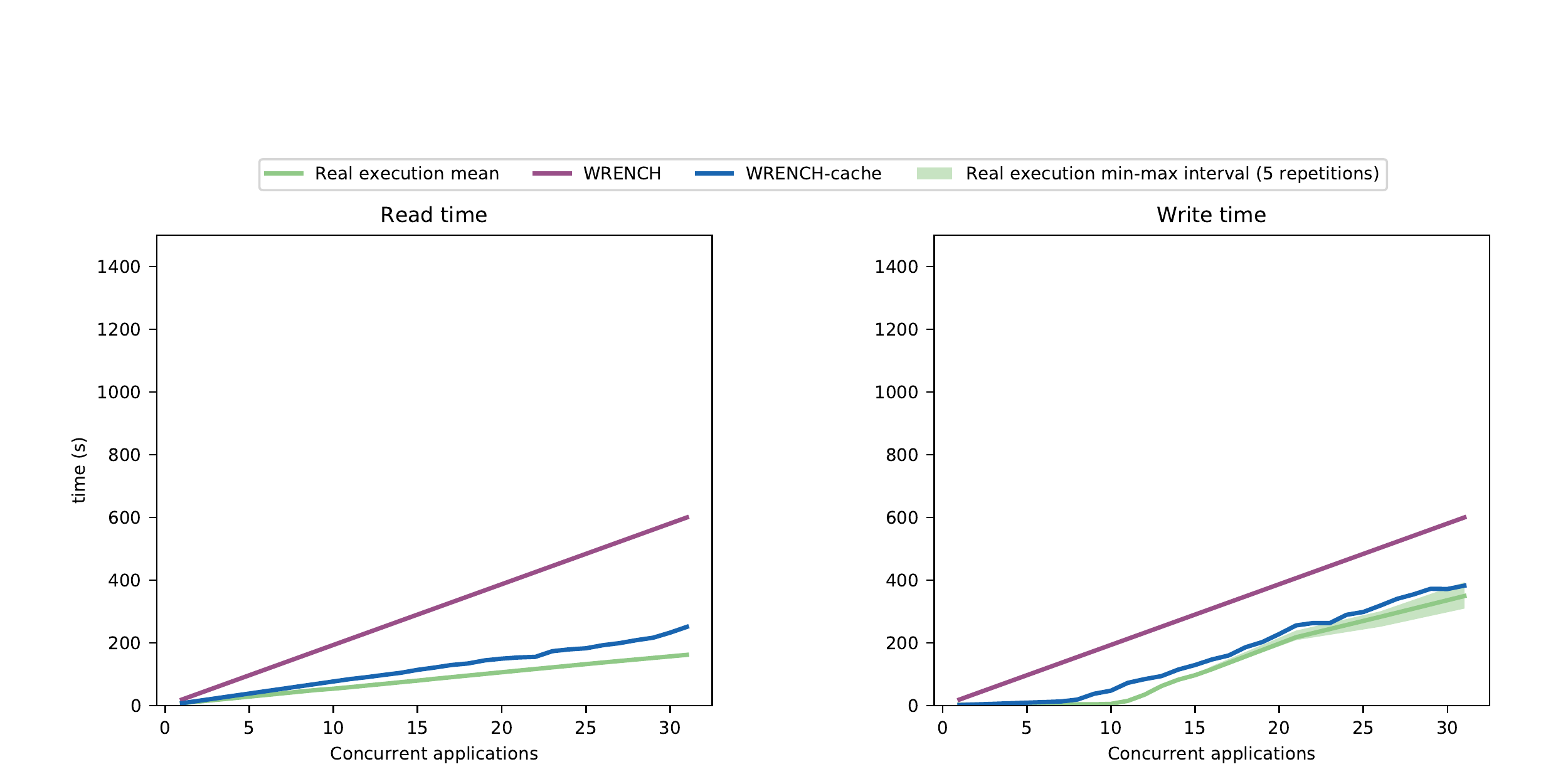}
                \end{subfigure}
                \caption{Concurrent results with 3~GB files (\textit{Exp 2})}
                \label{fig:multi_local}
            \end{figure*}
        \subsection{Concurrent applications (Exp 2)}

            The page cache model notably reduced \wrench's simulation error
            for concurrent applications executed with local I/Os
            (Figure~\ref{fig:multi_local}). For reads, \wrench-cache
            slightly overestimated runtime, due to the discrepancy between
            simulated and real read bandwidths mentioned before. 
            For writes, \wrench-cache
            retrieved a plateau similar to the one observed in the real
            execution, marking the limit beyond which the page cache was
            saturated with dirty data and needed flushing.

            \begin{figure}[b]
                \begin{subfigure}{0.95\linewidth}
                    \centering
                    \includegraphics[width=\linewidth]{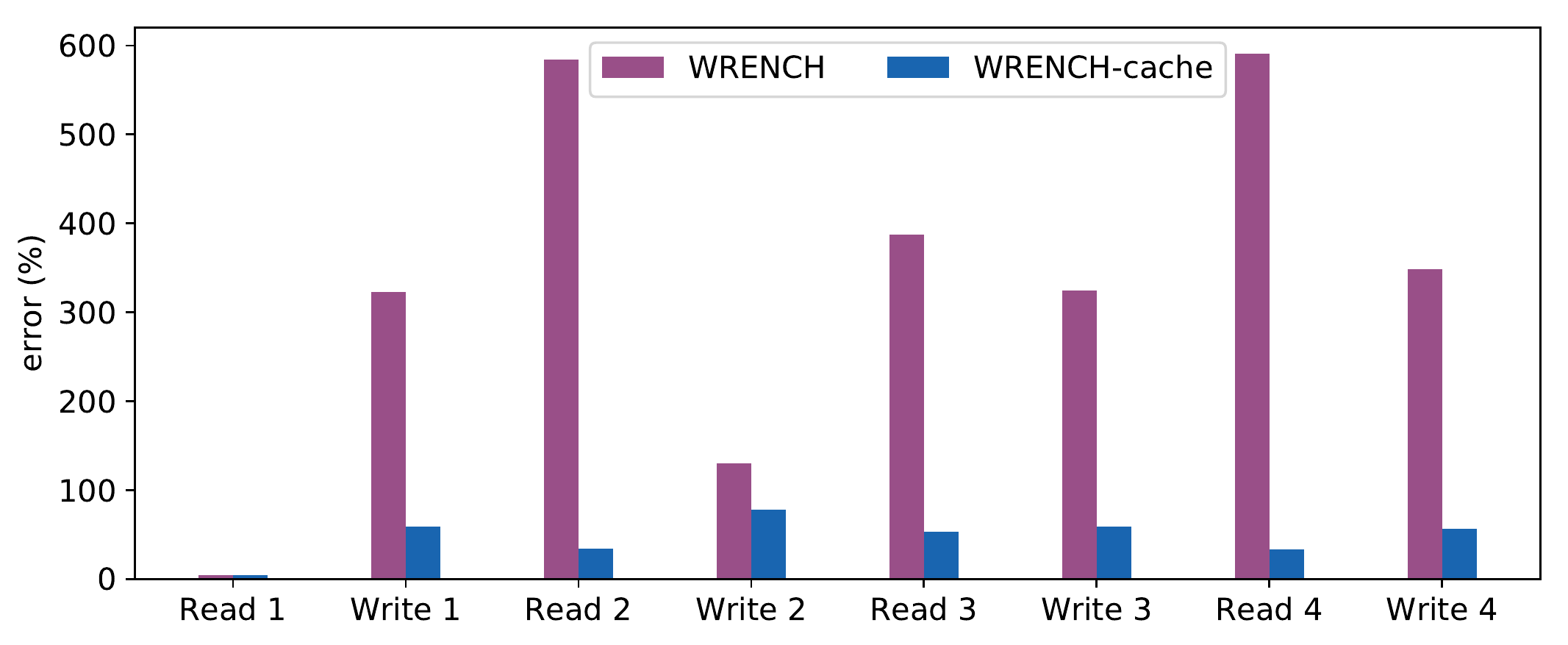}
                \end{subfigure}
                \caption{Real application results (\textit{Exp 4})}
                \label{fig:nighres}
                \end{figure}
        \subsection{Remote storage (Exp 3)}

\begin{figure*}
                \begin{subfigure}{\linewidth}
                    \centering
                    \includegraphics[width=\linewidth]{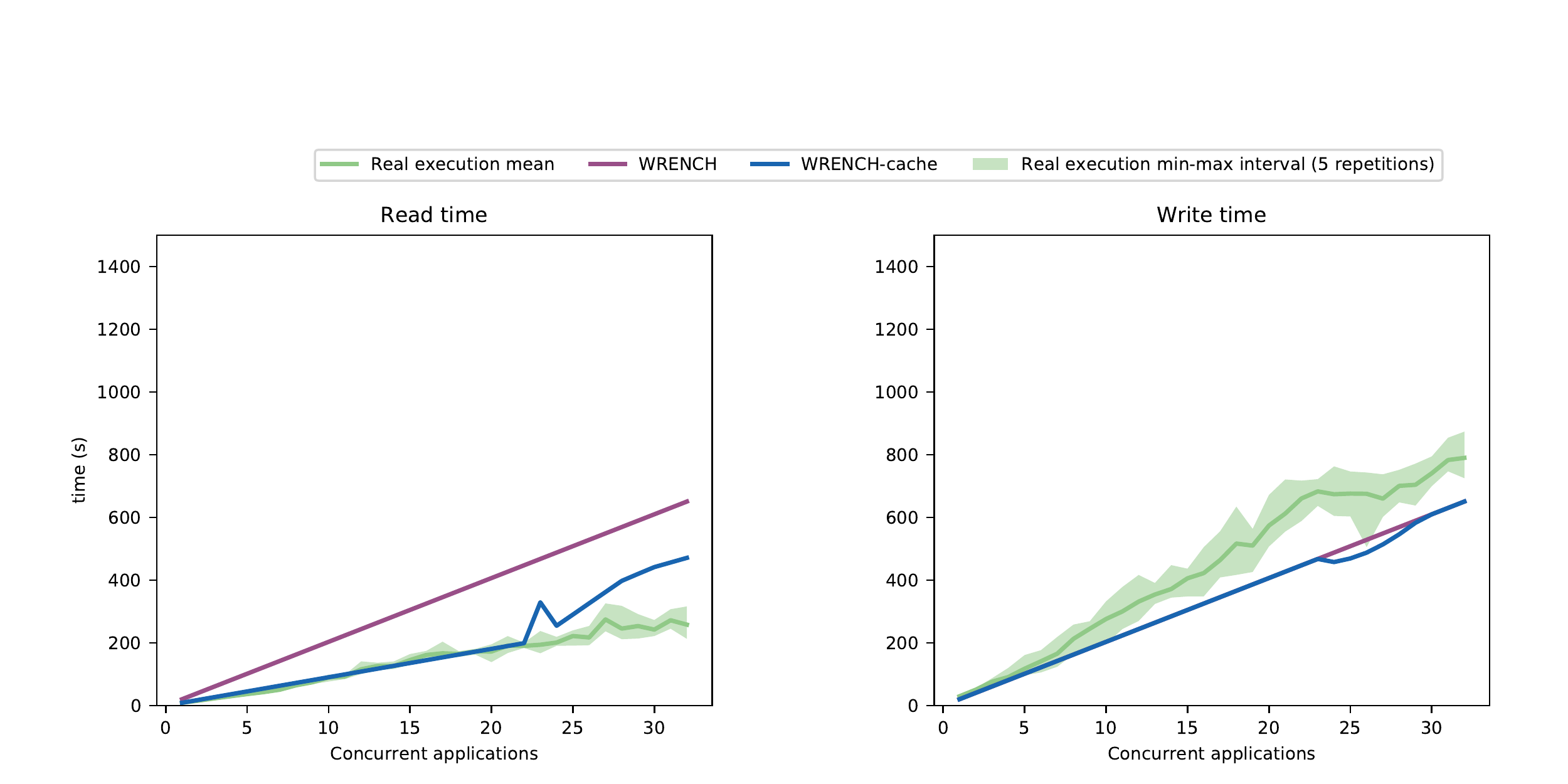}
                \end{subfigure}
                \caption{NFS results with 3~GB files (\textit{Exp 3})}
                \label{fig:multi_nfs}
                \end{figure*}

            Page cache simulation importantly reduced simulation error
            on NFS storage as well (Figure~\ref{fig:multi_nfs}). This
            manifested only for reads, as the NFS server used writethrough rather than writeback cache.
            Both \wrench and \wrench-cache
            underestimated write times due to the discrepancy between
            simulated and real bandwidths mentioned previously. For reads,
            this discrepancy only impacted the results beyond 22
            applications since before this threshold, most reads resulted in cache
            hits.

        \subsection{Real application (Exp 4)}
        Similar to the synthetic application, simulation errors were
        substantially reduced by the WRENCH-cache simulator compared to
        WRENCH (Figure~\ref{fig:nighres}). On average, errors were reduced
        from 337~\% in WRENCH to 47~\% in WRENCH-cache. 
        The first read happened entirely from disk and was therefore 
        very accurately simulated by both WRENCH and WRENCH-cache.

            \begin{figure}
                \begin{subfigure}{\columnwidth}
                    \centering
                    \includegraphics[width=\linewidth]{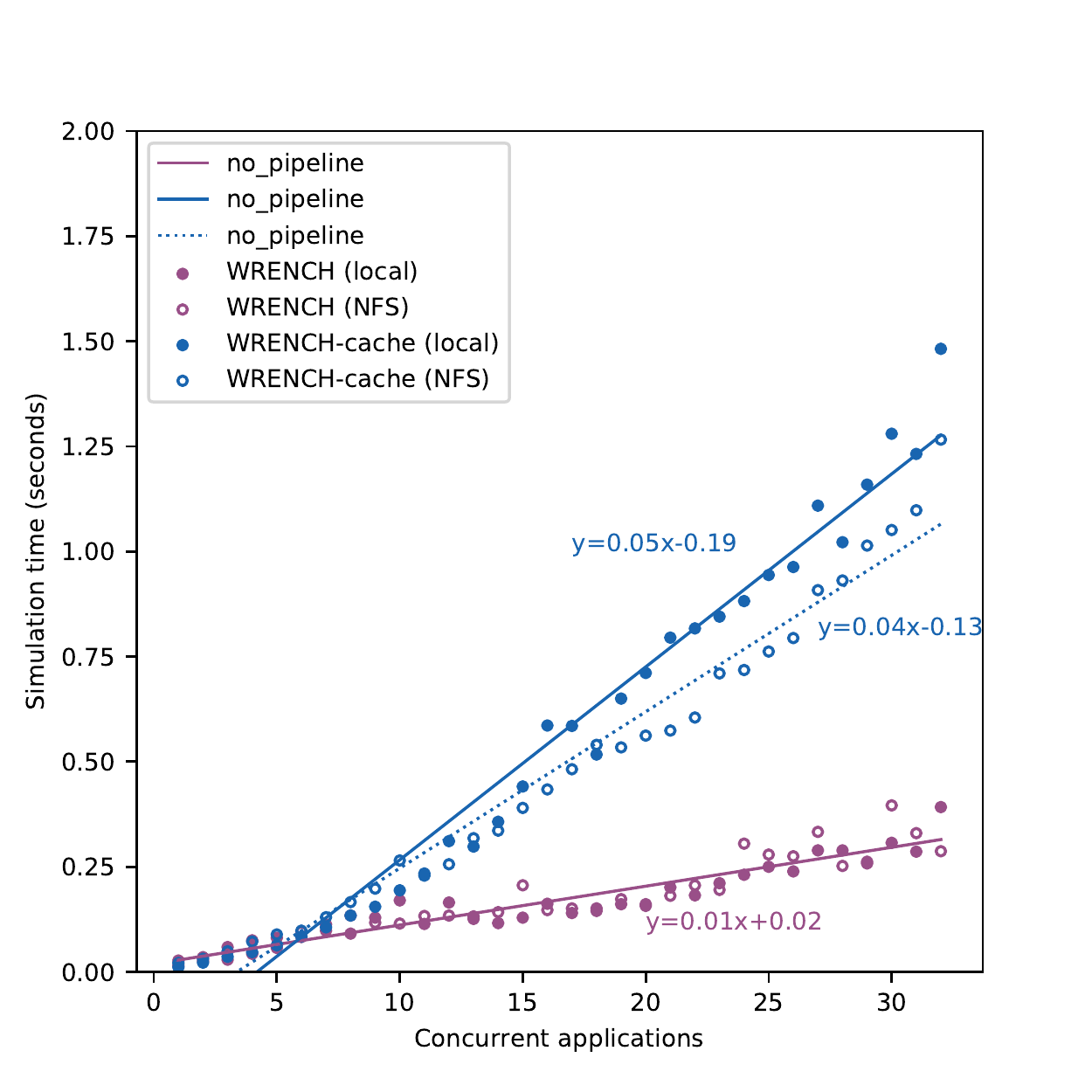}
                \end{subfigure}
                \caption{Simulation time comparison. \wrench-cache scales
                linearly with the number of concurrent applications, albeit
                with a higher overhead than \wrench.}
                \label{fig:multi_time}
                \end{figure}
    
        \subsection{Simulation time}
        As is the case for \wrench, simulation time with \wrench-cache scales
        linearly with the number of concurrent applications
        (Figure~\ref{fig:multi_time}, p \textless $10^{-24}$). However, the page
        cache model substantially increases simulation time by
        application, as can be seen by comparing regression slopes in
        Figure~\ref{fig:multi_time}. Interestingly, \wrench-cache is faster with 
        NFS I/Os than with local I/Os, most likely due to the use of writethrough
        cache in NFS, which bypasses flushing operations.

    \section{Conclusion}
    \label{discussion}
    We designed a model of the Linux page cache and implemented it in the
    \simgrid-based \wrench simulation framework to simulate the execution
    of distributed applications.
    Evaluation results show that our model improves simulation accuracy
    substantially, reducing absolute relative simulation errors by up to
    9$\times$ (see results of the single-threaded experiment). The
    availability of asymmetrical disk bandwidths in the forthcoming
    \simgrid release will further improve these results.
    Our page cache model is publicly available in the \wrench GitHub
    repository.

    Page cache simulation can be instrumental in a number of studies. For
    instance, it is now common for HPC clusters to run applications in
    Linux control groups (cgroups), where resource consumption is limited,
    including memory and therefore page cache usage. Using our simulator,
    it would be possible to study the interaction between memory allocation
    and I/O performance, for instance to improve scheduling algorithms or
    avoid page cache starvation~\cite{zhuang2017}. Our simulator could also
    be leveraged to evaluate solutions that reduce the impact of network
    file transfers on distributed applications, such as burst
    buffers~\cite{ferreiradasilva-fgcs-bb-2019}, hierarchical file
    systems~\cite{islam2015triple}, active storage~\cite{5496981}, or
    specific hardware architectures~\cite{hayot2020performance}. 

    Not all I/O behaviors are captured by currently available simulation models,
    including the one developed in this work, 
    which could substantially limit the accuracy of simulations.
    Relevant extensions to this work include more
    accurate descriptions of anonymous memory usage in applications, 
    which strongly affects I/O times through writeback cache. File access patterns
     might also be worth including in the simulation models,
    as they directly affect page cache content.
    % readahead and persistent storage could also be adeed
    % Simulation results could be made even more accurate by a deeper
    % investigation of page cache flushing and eviction order. 

        \section{Acknowledgments}
The computing platform used in the experiments was obtained with funding
from the Canada Foundation for Innovation. This work was partially supported
by NSF contracts \#1923539 and \#1923621.

\bibliographystyle{IEEEtran}
\bibliography{citation}

\end{document}